\documentclass[aps,showpacs,manuscript,12pt]{revtex4}
\usepackage{amsfonts}
\usepackage{amssymb}
\usepackage{amsmath}
\usepackage{graphicx}

\begin{document}

\title{\textbf{The exact radiation-reaction equation\\
for a classical charged particle$^{\S }$ }}
\author{M. Tessarotto$^{a,b}$, M. Dorigo$^{c}$, C. Cremaschini$^{d}$, P.
Nicolini$^{a,b}$ and A. Beklemishev$^{e}$}
\affiliation{$^{a}$Department of Mathematics and Informatics, University of Trieste,
Italy, $^{b}$Consortium of Magneto-fluid-dynamics, University of
Trieste,Italy, $^{c}$Department of Physics, University of Trieste, Italy, $%
^{d}$Department of Astronomy, University of Trieste, Italy, $^{e}$Budker
Institute of Nuclear Physics, Novosibirsk, Russia }

\begin{abstract}
An unsolved problem of classical mechanics and classical electrodynamics is
the search of the exact relativistic equations of motion for a classical
charged point-particle subject to the force produced by the action of its EM
self-field. The problem is related to the conjecture that for a classical
charged point-particle there should exist a relativistic equation of motion
(RR equation) which results both \textit{non-perturbative}, in the sense
that it does not rely on a perturbative expansion on the electromagnetic
field generated by the charged particle and \textit{non-asymptotic}, i.e.,
it does not depend on any infinitesimal parameter. In this paper we intend
to propose a novel solution to this well known problem, and in particular to
point out that the RR equation is necessarily variational. The approach is
based on two key elements: 1) the adoption of the relativistic hybrid
synchronous Hamilton variational principle recently pointed out (Tessarotto
et al, 2006). Its basic feature is that it can be expressed in principle in
terms of arbitrary "hybrid" variables (i.e., generally non-Lagrangian and
non-Hamiltonian variables); 2) the variational treatment of the EM
self-field, taking into account the exact particle dynamics.
\end{abstract}

\date{\today }
\maketitle



\section{Introduction}

A famous (and unsolved) theoretical issue both in classical and quantum
mechanics is related to the \emph{radiation reaction} (\emph{RR}) \emph{%
problem}, i.e., the treatment of the dynamics of a charged particle in the
presence of its EM self-field (for an introduction and background see
Feyman, 1970 \cite{Feynman1988}) to be based on the construction of its
relativistic RR equations of motion\textit{\ (\emph{RR equation})}. For
contemporary science the search of a possible \textit{exact} solution of the
RR problem represents not merely an unsolved intellectual challenge, but a
fundamental prerequisite for the proper formulation of all physical theories
which are based on the description of relativistic dynamics of classical
charged particles. These involve the consistent formulation of the
relativistic kinetic theory of charged particles and of the related fluid
descriptions (i.e., the relativistic magnetohydrodynamic equations obtained
by means of suitable closure conditions), both essential in plasma physics
and astrophysics. Despite efforts spent by many the problem of its
theoretical description remains still elusive. \

In classical mechanics the RR problem was first posed by Lorentz in his
historical work (Lorentz, 1985 \cite{Lorentz}; see also Abraham, 1905 \cite%
{Abraham1905}). Traditional approaches are based either on the RR equation
due to Lorentz, Abraham and Dirac (first presented by Dirac in 1938 \cite%
{Dirac1938}), nowadays popularly known as the \emph{LAD equation, }or the
equation derived from it by Landau and Lifschitz \cite{LL} via a suitable
"reduction process", the so-called \emph{LL equation. \ }As recalled
elsewhere [see related discussion in \cite{Dorigo2008a} (Ref.A)] several
aspects of the RR problem - and of the LAD and LL equations - are yet to
find a satisfactory formulation/solution.\ Common feature of all previous
approaches is the adoption of an asymptotic expansion for the EM self-field
(or for the corresponding EM 4-potential), rather than the exact
representation of the same force-field. This, in turn, implies that such
methods permit to determine - at most - only an asymptotic approximation for
the (still elusive) exact equation of motion for a charged particle subject
to its own EM self-field (RR equation). A side consequence of such
approximations is the possible violation of basic principles of classical
dynamics (for a review see for example \cite{Rohrlich1965} and related
discussion in Ref.A).

A major critical aspect of the RR problem is, however, related to its (still
missing) possible variational formulation. \ This is reflected by the
circumstance that \ - as pointed out in Ref.A\ - all RR equations obtained
so far (in particular the LAD and LL equations) are \textit{non-variational,}
i.e., they cannot be derived from a variational principle. \ This result is
clearly in contrast to the basic principles of classical mechanics. In
particular it conflicts with Hamilton's action principle, which - under such
premises (i.e., the validity of LAD and/or LL equations) - should actually
hold true only in the case of inertial motion! A major consequence which
follows is that the dynamics of point-like charged particles described by
these approximate model equations is not Hamiltonian. This is actually the
reason why in contemporary literature relativistic systems of charged
particles are not considered as Hamiltonian systems. Nevertheless, it is not
clear whether this feature is only an accident, i.e., is only due to the
approximations introduced in the RR equations adopted so far or is actually
intrinsic to the nature of the RR problem. Unfortunately, a satisfactory
answer to this fundamental question has not yet been given. Another key
issue is, however, related to the condition of validity of the relativistic
Hamilton variational principle \cite{Goldstein}.

In this paper we intend to analyze in detail a result which is already
well-known in the literature, namely that in its customary form the Hamilton
principle does not apply for point-particles. This is due to intrinsic
divergences (in particular due to the occurrence of an infinite EM mass)
produced by the EM self-field \cite{Feynman1988-b}: as a consequence, for
point-particles the radiation-reaction effect cannot be consistently taken
into account in the framework of classical electrodynamics. For this reason
in the past several authors, including Born and Infeld, Dirac, Wheeler and
Feynman (see discussion in Ref.\cite{Feynman1988}), tried to modify
classical electrodynamics in an effort to eliminate all divergent
contributions arising due to EM self-interactions. This is the so-called
regularization problem for point-particles, based on the introduction of
suitable modifications of Maxwell's electrodynamics. There is an extensive
literature devoted to possible ways to achieve this goal (for a review and
references on the subject see for example \cite{Quinn2001}). A possible
strategy involves introducing appropriate modifications of the EM self
4-potential. Typically this is done\ (see for example Rohlich \cite%
{Rohlich1964}) by assuming that there exists a decomposition of the EM
field, whereby each particle "feels" only the action of external particles
and of a suitable part of the EM self-field. While this decomposition
becomes clearly questionable for finite-size particles, its consistency with
first principles - and in particular with standard quantum mechanics - seems
dubious, to say the least. Indeed, according to Feynman's own's words \cite%
{Feynman1988} up to now "nobody ever succeeded in making a self-consistent
quantum theory out of any of the(se) modified theories".

In our view these motivations clearly indicate that the route to the
solution of the RR problem should be based on the search of the exact
relativistic RR equation, i.e., the construction of a \textit{%
non-perturbative RR equation}. In this paper we intend to propose a novel
solution to the RR problem, by pointing out that it can be achieved by means
of the relativistic Hamilton variational principle formulated in the
framework of classical electrodynamics. The approach is based on the
adoption of a synchronous variational principle \cite{Cremaschini2006} and
for finite-size spherical-shell charges. As a consequence, the explicit
variational treatment of the retarded EM self-potential generated by the
same particles is made possible. Based on the construction of the
Euler-Lagrange equations stemming from the variational principle, the
\textit{exact} relativistic equations of motion for a charged finite-size
particle immersed in prescribed EM and gravitational fields can in principle
be obtained in this way.

\section{Variational description of classical point-particle relativistic
dynamics}

From the mathematical viewpoint, one of the corner-stones of classical
mechanics is the assumption that the coupled set of equations formed by the
particle dynamical equations and Maxwell's equations is variational \cite%
{LL,Goldstein}. In other words, both the particle state and the EM field in
which the particle is immersed are completely determined by means of a
suitable variational principle. In relativistic classical mechanics it is
well known that - \textit{consistent with Maxwell's classical electrodynamics%
} - this is realized by the Hamilton variational principle, to be formulated
in the framework of a fully covariant description. The choice of the
dynamical variables which define the particle state remains in principle
arbitrary. Thus, they can always be represented by so-called "hybrid"
variables, i.e., superabundant variables which generally do not define a
Lagrangian state. This implies, thanks to Darboux theorem, that it should
always be possible to identify them locally with canonical variables. As a
basic consequence, classical systems should be necessarily Hamiltonian,
i.e., their canonical states should be extrema of the corresponding
Hamiltonian action, while the corresponding particle equations of motion,
i.e., the Euler-Lagrange equations provided by the same variational
principle, necessarily should coincide with Hamilton's equations of motion.
Here, in particular, we intend to show that the same variational principle
should hold also when: (a) the EM field is considered variational, namely
the variation of the action with respect to the EM 4-potential delivers also
the complete set of Maxwell's equations; (b) more generally, when the EM
field, specified via its 4-potential ($A_{\mu }$), is represented in terms
of an arbitrary "admissible" superposition of prescribed and variational
parts. By definition in the sequel a decomposition of the four potential of
the form $A_{\mu }=A_{\mu }^{(1)}+A_{\mu }^{(2)},$ \ where $A_{\mu }^{(1)}$
and $A_{\mu }^{(2)}$ denote 'a priori' arbitrary contributions to $A_{\mu },$
is denoted as \textit{admissible }for a prescribed variational functional if
all related contributions, appearing in the same functional, actually exist.
This means, that the considered field decomposition must be specified in
such a way that both $A_{\mu }^{(1)}$ and $A_{\mu }^{(2)}$ are defined in
the whole phase-space and are summable in a suitable sense so that all
relevant integrals involving the two 4-vectors actually can be uniquely
defined. Thus, if Hamilton variational principle holds for an arbitrary
choice of the EM field and arbitrary initial conditions for a system of
charged classical particles, it is expected to apply also in the presence of
the EM self-field generated by the particles themselves. \emph{In other
words, the relativistic equations of motion for charged classical particles
should be variational also}\textit{\ }\emph{in the presence of the radiation
reaction (EM self-force) arising from the particles themselves. }

In the following we shall stick purely to classical electrodynamics. For
this purpose we intend to adopt the classical Hamilton variational principle
for relativistic particles \cite{Goldstein}. For definiteness, let us first
consider the case of a charged point-particle immersed in an EM field,
identifying with $\left( r^{\mu },u_{\mu }\right) $ the particle state, with
$r^{\mu }$ and $u_{\mu }$ the position and velocity 4-vectors, and with $%
A_{\mu }(r)$ the EM 4-vector potential associated to the EM field (and
depending on the 4-vector $r\equiv r^{\nu }$). In classical mechanics the
variational functional (the Hamilton action functional) is well-known, and
can be realized either by means of asynchronous \cite{LL} or synchronous
\cite{Goldstein} variational principles. The variational functional (action
functional) is defined in terms of the curves $r^{\mu }(s)$ and $u_{\mu }(s),
$ functions of the proper time $s$ [with $s\in
\mathbb{R}
$], and the EM 4-potential $A_{\mu }(r)$. In the case of a point-particle
the following Theorem/Axiom should hold \cite{Cremaschini2006}

\textbf{THM.1 - Hamilton principle for point-particles } \emph{Let us assume
that:} \emph{1) the real functions }$f(s)\equiv \left[ r^{\mu }(s),u_{\mu
}(s),\chi (s)\right] $\emph{\ and the real 4-vector }$A_{\mu }(r)$\emph{\
belong respectively to suitable functional classes }$\left\{ f\right\} $
\emph{and }$\left\{ A_{\mu }\right\} $ \emph{in which end points and
boundaries are kept fixed; \ 2) the functional}%
\begin{eqnarray}
&&\left. S(r^{\mu },u_{\mu },\chi ,A_{\mu })=\int_{1}^{2}\left( m_{o}cu_{\mu
}+\frac{q}{c}A_{\mu }(r)\right) dr^{\mu }+\right.   \label{1} \\
&&+\int_{s_{1}}^{s_{2}}ds\chi (s)\left[ u_{\mu }(s)u^{\mu }(s)-1\right] +%
\frac{1}{16\pi c}\int \frac{d\Omega }{\sqrt{-g}}F^{\mu \nu }F_{\mu \nu }
\notag
\end{eqnarray}%
\emph{(Hamilton action integral) exists for all }$f(s)\in \left\{ f\right\} ,
$\emph{\ and }$A_{\mu }(r)\in \left\{ A_{\mu }\right\} .$ \emph{Here, }$%
u^{\mu }(s)=g^{\mu \nu }u_{\nu }(s),$\emph{\ while }$g^{\mu \nu }=g^{\mu \nu
}(r(s))$\emph{\ denotes the counter-variant components of the metric tensor,
each one to be considered dependent of the generic varied curve }$r(s);$%
\emph{\ furthermore, }$F_{\mu \nu }=\partial _{\mu }A_{\nu }-\partial _{\nu
}A_{\mu }$\emph{\ and }$F^{\mu \nu }=g^{\mu \alpha }g^{\nu \beta }F_{\alpha
\beta },$\emph{\ while }$m_{o}$\emph{\ and }$q$\emph{\ are respectively the
constant rest mass and electric charge of a point particle, }$d\Omega =\sqrt{%
-g}dtdxdydz$\emph{\ the 4-volume element and }$ds$\emph{\ the line element;
3) if }$f(s),A_{\mu }(r)$ \emph{are extremal curves of }$S$ \emph{(see below)%
} \emph{the line element }$ds$\emph{\ satisfies the constraint }$%
ds^{2}=g_{\mu \nu }(r(s))dr^{\mu }(s)dr^{\nu }(s).$ \emph{It follows that : T%
}$_{1}$\emph{) for arbitrary independent synchronous variations }$\delta
f(s),$\ \emph{the synchronous variational principle }%
\begin{equation}
\delta S=0,  \label{variational principle}
\end{equation}%
\emph{delivers the following set of Euler-Lagrange equations for the
extremal curves }$f(s)$\emph{:}%
\begin{eqnarray}
&&\left. -d\left( m_{o}cu_{\mu }+\frac{q}{c}A_{\mu }\right) +\frac{q}{c}%
\frac{\partial }{\partial r^{\mu }}A_{\nu }dr^{\nu }+2u_{\alpha }u_{\beta
}\chi (s)\partial _{\mu }\left( g^{\alpha \beta }\right) =0,\right.
\label{3a-1} \\
&&\left. m_{o}cdr^{\mu }+2\chi (s)u^{\mu }(s)ds=0,\right.   \label{3a-2} \\
&&\left. u_{\mu }(s)u^{\mu }(s)-1=0,\right.   \label{3a-3}
\end{eqnarray}%
\emph{\ where the extremal value of the Lagrange multiplier }$\chi $\emph{\
reads for all }$s\in
\mathbb{R}
,$ $\chi (s)=-\frac{m_{o}c}{2};$ \emph{T}$_{2}$\emph{) for arbitrary
variations }$\delta A_{\mu }$ \emph{in which }$\delta A_{\mu }$\emph{\ is
considered independent of the extremal curve }$r^{\nu }(s),$ \emph{Eq.(\ref%
{variational principle}) delivers as Euler-Lagrange equation the Maxwell's
equations for the EM 4-potential }$A_{\mu }(r):$\emph{\ }%
\begin{equation}
\left. \partial _{\mu }F^{\mu \nu }=\frac{4\pi }{c}j^{\nu },\right.
\label{3a-4}
\end{equation}%
\emph{where the 4-vector }$j^{\mu }=j^{\mu }(r)$\emph{\ in Eq.(\ref{3a-4})
reads }$j^{\mu }(r)=qc\int_{s_{1}}^{s_{2}}ds^{\prime }u^{\mu }(s^{\prime
})\delta ^{(4)}\left( r-r(s^{\prime })\right) ,$ \emph{i.e., it is the
4-current of a point charge moving along the world-line }$r^{\mu }=r^{\mu
}(s)$\emph{\ with a 4-velocity }$u^{\mu }(s).$

PROOF

The proof of $T_{1}$ is straightforward (see Ref.\cite{Cremaschini2006}). To
obtain the Euler-Lagrange equation for $\delta A_{\mu }$ one invokes the
identities
\begin{equation}
\frac{1}{16\pi }\delta \int \frac{d\Omega }{\sqrt{-g}}F^{\mu \nu }F_{\mu \nu
}=-\frac{1}{4\pi }\int \frac{d\Omega }{\sqrt{-g}}\delta A_{\nu }\partial
_{\mu }F^{\mu \nu },
\end{equation}%
\begin{equation}
\delta \int_{s_{1}}^{s_{2}}dsA_{\mu }(r(s))\frac{dr^{\mu }(s)}{ds}=\int
\frac{d\Omega }{\sqrt{-g}}\int_{s_{1}}^{s_{2}}ds^{\prime }\delta A_{\mu
}(r(s^{\prime }))\frac{dr^{\mu }(s^{\prime })}{ds^{\prime }}\delta
^{(4)}\left( r(s)-r(s^{\prime })\right) ,
\end{equation}%
where $\delta ^{(4)}\left( r-r(\tau )\right) $ is the 4-dimensional Dirac
delta. \ Requiring that $\delta A_{\mu }$ is independent of the 4-vector $%
r^{\nu }(s)$ the variational principle (\ref{variational principle})
delivers the Maxwell's equations (\ref{3a-4}), in which the 4-current $%
j^{\mu }(r^{\nu })$ necessarily takes the form defined in THM.1.

Let us now investigate the conditions of validity of THM.1. It is important
to remark that if this theorem is true for an arbitrary (suitably smooth) EM
4-potential $A_{\mu }(r)$, it must apply also if the EM 4-potential $A_{\mu
}(r)$ is represented in terms on an arbitrary decomposition for the
4-potential. Thus for example, one of the two components (say $A_{\mu
}^{(1)} $) can be in principle considered prescribed, in the sense that
there results by assumption $\delta A_{\mu }^{(1)}\equiv 0$ for all
synchronous variations such that $\delta f(s)\equiv 0$ (as an example, we
may consider the trivial decomposition in which also $A_{\mu }^{(2)}\equiv 0$%
)$.$ Therefore, in particular, the Hamilton variational principle [Eq.(\ref%
{variational principle})] should hold true also in the case in which the EM
4-potential takes the form
\begin{equation}
A_{\mu }=A_{\mu }^{(self)}+A_{\mu }^{(ext)},
\label{fundamental
decomposition}
\end{equation}%
i.e., it is represented in terms of the EM self- and external 4-potentials,
while letting $\delta A_{\mu }^{(self)}\equiv 0,$ for all variations $\delta
A_{\mu }=\delta A_{\mu }^{(ext)}$ and such that $\delta f(s)\equiv 0.$ Since
- thanks to the linearity of Maxwell's equations - this decomposition can
always be made, it should be admissible, by definition, for the variational
functional (\ref{1}). On the other hand, its possible violation\ (i.e., in
case the decomposition is not admissible) would have the fundamental
consequence that the Hamilton variational principle \textit{becomes invalid
for point-particles when the EM self-field of the point-particle is taken
into account}. In such a case the following result can be proven:

\textbf{THM.2 - Violation of THM.1 for the EM self-force of point-particles}

\emph{As a consequence of THM.1 it follows that: \ C}$_{1}$\emph{) the
Euler-Lagrange equations obtained by imposing an arbitrary synchronous
variation }$\delta A_{\mu }$\emph{\ must hold for any decomposition }$A_{\mu
}(r)=A_{\mu }^{(1)}(r)+A_{\mu }^{(2)}(r)$ \emph{of the 4-vector }$A_{\mu
}(r) $ \emph{which is admissible for (\ref{1})}$,$ \emph{where one of the
two terms, for example }$A_{\mu }^{(1)}(r),$\emph{\ is considered a
prescribed function of the 4-vector }$r(s)\in \left\{ f\right\} ,$ \emph{%
i.e., such that }$\delta A_{\mu }^{(1)}(r)=0$\emph{\ when }$\delta
f(s)\equiv 0;$ \emph{C}$_{2}$\emph{) the decomposition (\ref{fundamental
decomposition})\ is not admissible for the action (\ref{1}), i.e., in terms
of the 4-potentials of the EM self-field and of the external EM field}, $%
A_{\mu }^{(self)}$\emph{\ and }$A_{\mu }^{(ext)}$.

PROOF

The proof of proposition \emph{C}$_{1}$\emph{\ }is as follows. Let us
consider and arbitrary admissible decomposition $A_{\mu }(r)=A_{\mu
}^{(1)}(r)+A_{\mu }^{(2)}(r),$ requiring that\emph{\ }$A_{\mu }^{(1)}(r)$ is
a prescribed function of the extremal curve $r^{\nu }(s),$\ so that there
results identically $\delta A_{\mu }\equiv \delta A_{\mu }^{(2)}.$ The
variational equation for $\delta A_{\mu }$ remains manifestly unchanged. To
prove proposition \emph{C}$_{2},$ let us now pose the problem whether Eq. (%
\ref{fundamental decomposition}) is an admissible decomposition or not. For
definiteness, let us consider the particular case of flat space-time (i.e., $%
\sqrt{-g}=1$). \ In this case the expression of $A_{\mu }^{(self)}$ for
point-particles\ is well-known and coincides with the so-called
Lienard-Wiechart potentials \cite{LL}. \ It is immediate to prove that $%
A_{\mu }^{(self)}$ carries diverging contributions to the action functional $%
S,$ in particular due to the integral $\int \frac{d\Omega }{\sqrt{-g}}%
F^{(self)\mu \nu }F_{\mu \nu }^{(self)}$ which is manifestly divergent.
Hence the decomposition (\ref{fundamental decomposition}) is not admissible.

\section{Variational description for finite-size charges}

THM.2 implies the fundamental consequence that for point-particles the
variational principle (\ref{variational principle}) becomes invalid if the
EM 4-potential $A_{\mu }^{(self)}$ is properly taken into account. This is
due to the divergences produced by the self-force generated by the point
particle. \ To deal with this basic difficulty several approaches have been
attempted in the past \cite{Detweiler,Quinn,Kosyakov} (for a summary see the
discussion in Ref. \cite{Gal'tsov2008}) by adopting various types of
axiomatic assumptions about the nature of the singular terms. Nevertheless,
their possible consistent derivation from first principles is still missing
\cite{Gal'tsov2008}. \ \emph{In the following we intend to show, however,
that the validity of Hamilton variational formulation can be restored -
without the introduction of any additional assumption - if the
point-particles are replaced by finite-size charges, defined in such a way
that the EM self-potential }$A_{\mu }^{(self)}$\emph{\ remains always
finite. }For definiteness, let us consider as in Ref.A a \emph{finite-size
spherical-shell charge} carrying constant rest mass and total charge,\emph{\
}$m_{o}$ and $q$. The particle charge is assumed to be uniformly distributed
on a spherical shell of finite radius $\sigma >0,$ which carries the
homogeneous surface charge density $\rho =q/4\pi \sigma ^{2}$ (defined with
respect to a frame locally at rest w.r. to the particle). In addition, if
the gravitation self-force is ignored, the particle mass may be treated as
concentrated in the center of the sphere so that the particle degree of
freedom is the same as that of a point particle. One can prove that\emph{\
in such a case the construction of an exact relativistic RR equation can be
achieved simply by identifying it with the appropriate Euler-Lagrange
equation determined by the Hamilton principle.} This is obtained by an
appropriate generalization of THM.1. It follows by introducing in the
previous definition of the action integral the formal replacements $%
ds\rightarrow W(r,s)\frac{d\Omega }{\sqrt{-g}}$ and $dr^{\mu }\rightarrow
\frac{dr^{\mu }}{ds}W(r,s)\frac{d\Omega }{\sqrt{-g}},$ where $W(r,s)$ (the
"wire function") is generally to be identified with a suitable distribution
and $r$ here denotes the 4-vector $r^{\mu }(s)$. For a generic wire-function
the functional (\ref{1}) becomes%
\begin{eqnarray}
&&\left. S(r^{\mu },u_{\mu },\chi ,A_{\mu })=\int \frac{d\Omega }{\sqrt{-g}}%
W(r,s)\left( m_{o}cu_{\mu }(s)+\frac{q}{c}A_{\mu }(r)\right) \frac{dr^{\mu
}(s)}{ds}+\right.   \label{2'} \\
&&+\frac{1}{16\pi c}\int \frac{d\Omega }{\sqrt{-g}}F^{\mu \nu }F_{\mu \nu
}+\int_{s_{1}}^{s_{2}}\frac{d\Omega }{\sqrt{-g}}W(r,s)\chi (s)\left[ u_{\mu
}(s)u^{\mu }(s)-1\right] ,  \notag
\end{eqnarray}%
while - similarly - the 4-current $j^{\mu }(r^{\nu })$ reads $j^{\mu
}(r^{\nu })=qc\int \frac{d\Omega ^{\prime }}{\sqrt{-g}}W(r^{\prime
},s^{\prime })u^{\mu }(s^{\prime })\delta ^{(4)}\left( r-r(s^{\prime
})\right) .$ In the case of a spherical-shell particle immersed in a
Minkowsky space-time, when expressed in a reference frame locally at rest
with respect to the particle (\emph{rest-frame}), the wire-function $W$ must
be identified with%
\begin{equation}
W=\frac{\sqrt{-g}}{4\pi \sigma ^{2}}\delta \left( \left\vert \mathbf{r}-%
\mathbf{r}(s)\right\vert -\sigma \right) ,  \label{wire-function}
\end{equation}%
where $\sqrt{-g}=1.$ Hence, in the rest-frame there results $\int \frac{%
d\Omega }{\sqrt{-g}}W(r,s)\equiv \frac{1}{4\pi \sigma ^{2}}%
\int\limits_{s_{1}}^{s_{2}}ds\int\limits_{0}^{\infty }d\rho \rho ^{2}\int
d\Sigma (\mathbf{n})\delta \left( \rho -\sigma \right) =\frac{1}{4\pi }%
\int\limits_{s_{1}}^{s_{2}}ds\int d\Sigma (\mathbf{n}).$ Here all quantities
are evaluated in the rest-frame, hence $ds=cdt,$ while $d\Sigma (\mathbf{n})$
is the solid angle and $\mathbf{n}$ is the normal unit 3-vector on the unit
sphere. It follows that the appropriate generalization of Hamilton's
variational principle (again to be expressed in synchronous form \cite%
{Cremaschini2006}) requires that both $f\equiv \left[ r^{\mu },u_{\mu },\chi %
\right] $ and the 4-vector $A_{\mu }$ must generally be considered as
functions\ of $(\mathbf{n,}$ $s).$ Then the following theorem has the flavor
of (for further details we refer to Ref.\cite{Tessarotto2008c}):

\textbf{THM.3 - Hamilton principle for finite-size spherical-shell charges }

\emph{The action functional appropriate in case of the wire-function (\ref%
{wire-function}) is taken of the form }%
\begin{eqnarray}
&&\left. S(r^{\mu },u_{\mu },\chi ,A_{\mu })=\frac{1}{4\pi }%
\int_{s_{1}}^{s_{2}}ds\int d\Sigma (\mathbf{n})\left( m_{o}cu_{\mu }(\mathbf{%
n,}\text{ }s)+\frac{q}{c}A_{\mu }(r(\mathbf{n,}\text{ }s))\right) \frac{%
dr^{\mu }(\mathbf{n,}\text{ }s)}{ds}+\right.   \label{2"} \\
&&+\frac{1}{16\pi c}\int \frac{d\Omega }{\sqrt{-g}}F^{\mu \nu }F_{\mu \nu }+%
\frac{1}{4\pi }\int_{s_{1}}^{s_{2}}ds\int d\Sigma \text{ }\chi (\mathbf{n,}%
\text{ }s)\left[ g^{\mu \nu }(r(\mathbf{n,}\text{ }s))u_{\mu }(\mathbf{n,}%
\text{ }s)u_{\nu }(\mathbf{n,}\text{ }s)-1\right] ,  \notag
\end{eqnarray}%
\emph{where the line element }$ds$\emph{\ is required to satisfy the
constraint \#3 of THM.1. It follows that: T}$_{1}$\emph{) The Euler-Lagrange
equations obtained considering as independent the synchronous variations }$%
\delta f(\mathbf{n},s)$\emph{\ and }$\delta A_{\mu }(\mathbf{n},s)$\emph{\
yield identically Eqs.(\ref{3a-1}),(\ref{3a-2}),(\ref{3a-3}) and (\ref{3a-4}%
); T}$_{2}$\emph{) \ the Euler-Lagrange equations obtained by imposing an
arbitrary synchronous variation }$\delta A_{\mu }$\emph{\ hold for any
decomposition }$A_{\mu }(r)=A_{\mu }^{(1)}(r)+A_{\mu }^{(2)}(r)$ \emph{of
the 4-vector }$A_{\mu }(r)$ \emph{which is admissible for (\ref{2"}). In
particular, the decomposition (\ref{fundamental decomposition}) is
admissible for the action (\ref{1}), where }$A_{\mu }^{(self)}$\emph{\ and }$%
A_{\mu }^{(ext)}$\emph{\ are respectively the 4-potentials of the EM
self-field and of the external EM field.}

PROOF

The proof \ of proposition $T_{1}$ is similar to that given in THM.1. In
particular, the Euler-Lagrange equations for $\delta A_{\mu }(\mathbf{n,}s)$
- again to be identified with Maxwell's equations - follow by noting that
the functional $\int ds\int d\Sigma (\mathbf{n})A_{\mu }(r(s))\frac{dr^{\mu
}(\mathbf{n,}\text{ }s)}{ds}$ can also be written as $\int ds^{\prime }\int
\frac{d\Omega ^{\prime }}{\sqrt{-g}}\int d\Sigma (\mathbf{n})A_{\mu
}(r(s^{\prime }))\frac{dr^{\mu }(\mathbf{n,}\text{ }s^{\prime })}{ds^{\prime
}}\delta ^{(4)}\left( r-r(s^{\prime })\right) $. Hence, the 4-current $%
j^{\mu }(r)$ now reads necessarily
\begin{equation}
j^{\mu }(r)=\frac{qc}{4\pi }\int ds^{\prime }\int d\Sigma (\mathbf{n})\frac{%
dr^{\mu }(\mathbf{n,}\text{ }s^{\prime })}{ds^{\prime }}\delta ^{(4)}\left(
r-r(s^{\prime })\right) .
\end{equation}%
For a detailed proof of THM.3 (in particular of proposition $T_{2}$), which
requires the explicit calculation of the retarded EM 4-potential $A_{\mu
}^{(self)},$ we refer to \cite{Tessarotto2008c}. The main consequence of
THM.3 is that the Euler-Lagrange equation obtained by means of the
synchronous variation of the Hamilton action functional (\ref{2"}) with
respect to $\delta r^{\mu }(\mathbf{n},s)$ yields a possible realization of
the \emph{exact relativistic RR\ equation of motion }for a point-particle
carrying a finite-size classical charge.

\section{Conclusions}

In this paper the variational treatment of the radiation-reaction problem
has been investigated. First we have analyzed the Hamilton variational
principle, proving that it becomes invalid for charged point-particles if
the proper form of the EM self-field prescribed by classical electrodynamics
is taken into account. This conclusion (contained in THM. 1 and 2), which is
consistent with the customary interpretation of the RR problem \cite%
{Rohrlich1965}, is of course scarcely surprising. In fact, it is well-known
that the RR equation of motion becomes invalid for point-particles (see
Ref.A). \textit{However - as proven in this paper - the validity of Hamilton
principle can be restored for finite-size charges which admit a finite EM
self-potential} $A_{\mu }^{(self)}.$ In particular, the result can be
achieved by considering finite-size spherical-shell charges (THM.3) (this
choice is consistent with the approach used in Ref.A to construct the LAD
equation). Remarkable consequences which can be drawn from these conclusions
include: 1) \textit{the treatment of the RR problem can be achieved via a
variational formulation (Hamilton variational principle); }2)\textit{\ the
variational formulation is made transparent by adopting a synchronous
variational principle; 3})\textit{\ for prescribed finite-size charges in
principle the exact relativistic RR equations of motion can be achieved in
this way} \cite{Tessarotto2008c}. This suggests that: A)\textit{\ it should
be possible to extend the validity of the theory to curved space-time} and
moreover that B) \textit{\ analogous conclusions should apply also for the
gravitational radiation-reaction problem, i.e., in other words, the complete
treatment of the full EM-gravitational self-force should be possible in the
framework of a variational formulation}. For these reasons, the present
results appear of primary importance for relativistic theories (such as the
kinetic theory of charged particles and the gyrokinetic theory for
magnetoplasmas in curved space-time \cite{Cremaschini2006}) and related
applications in astro- and plasma physics.



\section*{Acknowledgments}

Work developed in cooperation with the CMFD Team, Consortium for
Magneto-fluid-dynamics (Trieste University, Trieste, Italy). \ Research
developed in the framework of the MIUR (Italian Ministry of University and
Research) PRIN Programme: \textit{Modelli della teoria cinetica matematica
nello studio dei sistemi complessi nelle scienze applicate}. The support
(A.B) of ICTP (International Center for Theoretical Physics, Trieste,
Italy), (A.B.) University of Trieste, Italy, (M.T) COST Action P17 (EPM,
\textit{Electromagnetic Processing of Materials}) and (M.T. and P.N.) GNFM
(National Group of Mathematical Physics) of INDAM (Italian National
Institute for Advanced Mathematics) is acknowledged.

\section*{Notice}

$^{\S }$ contributed paper at RGD26 (Kyoto, Japan, July 2008). 

\end{document}